\documentclass{phbauth}
\usepackage{graphicx}

\begin{document}

\begin{frontmatter}

\title{Surface Influence on Flux Penetration into HTS Bulks}

\author[address1]{A. A. Kordyuk\thanksref{thank1}},
\author[address2]{G. Krabbes},
\author[address1]{V. V. Nemoshkalenko},
\author[address1]{R. V. Viznichenko},

\address[address1]{Institute of Metal Physics, Kyiv, Ukraine}

\address[address2]{Institute of Solid State and Materials Research Dresden,
Germany}

\thanks[thank1]{Corresponding author. Present address: Institute of Metal
Physics, 36 Vernadsky str., Kyiv 252680, Ukraine, E-mail: kord@imp.kiev.ua}

\begin{abstract}
The influence of surface treatment on AC loss in melt-processed quasi-single 
crystal HTS was investigated with resonance oscillations technique. 
We have found that amplitude dependencies of AC loss on magnetic field 
amplitude become rather complicated after surface polishing. 
The experimental data show well distinguished dynamic crossover from
absence of barrier at low rates of field variation to its appearence at
higher rates. An explaination of such a dynamic surface barrier appearance
based on consideration of along surface vortex propagation was suggested.
\end{abstract}

\end{frontmatter}

\section{Introduction}

In parper \cite{JAP} the approach to calculate mechanical properties of 
levitation systems with melt-processed high temperature superconductors 
(MP HTS) was introduced.
The approach starts from an `ideally hard superconductor' approximation which
assumes that the penetration depth $\delta$ of alternating magnetic field is
zero.
Within this `zero' approximation the stiffness or resonance frequencies in
the permanent magnet (PM)--MP HTS system can be calculated analytically and
appeared to be in a good agreement with the experiment \cite{Met},
but to calculate energy loss due to PM motion \cite{MSEB} or hysteresis of
levitation force \cite{APL2} the next (`first') approximation has to be used
and finite values of $\delta$ have to be considered.

In \cite{MSEB} we have shown, that the energy loss $W$ in the PM--HTS system
during PM oscillations is mostly determined by AC loss in the HTS undersurface
layer $\delta = (c/4\pi)h_r/J_\mathrm{c}$, where $h_r$ is the tangential 
component of AC field at the HTS surface $S$ and $J_\mathrm{c}$ is the 
critical current density in $ab$-plane for field parallel to this plane, 
and for initial MP samples can be subdivided into two parts: $W = \int_S dS 
\left(\alpha h_r^3 + \beta h_r^2 \right)$.
The first part is well known bulk hysteretic loss within critical state model
from which $J_\mathrm{c}$ \cite{MSEB} and even its profiles \cite{PhyC} can be
determined.
In this paper we consider the second part of $W$ and investigate the
effect of surface treatment (polishing) on $W(h)$ dependence.
We discuss a possibility for thermal activation through surface barrier 
to be detected here and introduce an idea of dynamic surface barrier
appearance.  

\section{Experiment and Discussion}

\begin{figure}[t]
\begin{center}\leavevmode
\includegraphics[width=1\linewidth]{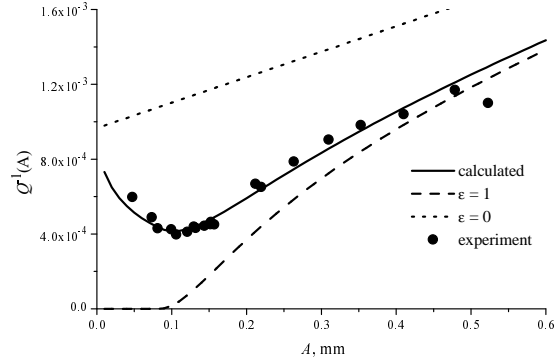}
\caption{
Inverse $Q$-factor vs. PM amplitude: experimental and calculated data.
}\label{fig1}\end{center}\end{figure}

Fig.\ \ref{fig1} represents the experimental dependencies of inverse
$Q$-factor of PM forced oscillations at resonance frequency $\omega$ on PM 
amplitude $A \propto h$.
$Q^{-1} = 2\pi W/W_0 \propto W/h^2$, where $W_0 \propto A^2$ is storage
energy. Symbols represent the data for the MP HTS sample with polished
top surface; dotted line shows $Q^{-1}(A)$ dependence for depolished sample. 

To explain the presence of the part of $W$ which is $\propto A^2$, a motion
of perpendicular to the surface vortices with an amplitude $s(A)$ was considered. 
In axially symmetrical configuration ${\bf r} = (r,z)$, due to small value of 
$\delta$, we can say that normal to the surface AC magnetic field component 
$b_z$ is determined by $h_r(r)$ distribution: 
$b_z = (1/2\pi r)(\mathrm{d}\Delta\Phi_r/\mathrm{d}r)$, where 
$\Delta\Phi_r = (c r/4 J_\mathrm{c})h_r^2$ is the parallel to surface magnetic 
flux variation.
This is true for $b_z \ll b_r$ which in our case is reinforced by anisotropy:
$J_\mathrm{c}({\bf B}\|c) \ll J_\mathrm{c}({\bf B}\|ab)$.
The function $s(r,A)$ can be obtained from the equation
\begin{eqnarray}
rB_z(r) - (r-s) B_z(r-s)=-r b_z(r,A), \label{1}
\end{eqnarray}
\noindent
where $B_z(r)$ is distribution of normal component of `frozen' magnetic field.
In such a way we have shown that for HTS with uniform bulk properties both 
parts of AC loss are related to vortex motion in HTS volume. 

By polishing the surface of sample we introduce a surface barrier for flux 
entry which causes a field jump $\Delta h (B_r)$ in undersurface `vortex free 
region' \cite{Clem1}. 
The influence of $\Delta h$ on $W$ can be taken into account by substitution
$h_r - \Delta h$ instead of $h_r$ and adding the surface loss as it was
made by Clem \cite{Clem2}. The dependence $Q^{-1}(A)$ which is obtained in
such a way is shown in Fig.\ \ref{fig1} as a dashed line. So, we can deduce
that experimental data show transition from absence of barrier at $A=0$ to
distinguish barrier at $A>0$, and
the first reason which comes to mind here is thermally activated flux 
penetration over the barrier \cite{Burl}, but it seems to be impossible
to describe the experimentally observed barrier disappearence at small $A$ 
using the relations from \cite{Burl}. 

This makes to suppose another possible mechanism to suppress the surface 
effect at low amplitudes. It is quite natural to expect that the barrier leads 
to $\Delta h$ not at whole surface but at its part $\varepsilon$ only. 
The reason for this is nonuniform flux penetration: when a part of vortex
or vortex bundle has penetrated into HTS, the further penetration can
take place without surmounting the barrier but with vortex propagation along 
the surface. 
Fixing vortex velocity
the part of vortices that
penetrates through barrier can be calculated by energy loss minimizing.
We have found: $\varepsilon = (1 + \zeta/h)^\mu$, where $\mu = -1/2$ and
$\zeta \approx 25$ Oe.
Then AC loss
\begin{eqnarray}
W(h)=\varepsilon W(h,\Delta h) + (1-\varepsilon)W(h,0). 
\label{2}
\end{eqnarray}
\noindent
The dependence (\ref{2}) is represented in Fig.\ \ref{fig1} by solid line.
Dashed line and dotted line represents the dependencies $W(h)$ for
$\varepsilon = 0$ and 1 respectively.

\begin{thebibliography}{9}

\bibitem{JAP}
A.A. Kordyuk, J. Appl. Phys. {\bf 83}, 610 (1998).
\bibitem{Met}
A.A. Kordyuk, Metal Phys. Adv. Tech. {\bf 18}, 249 (1999).
\bibitem{MSEB}
A.A. Kordyuk et al., Mat. Sci. Eng. B {\bf 53}, 174 (1998).
\bibitem{APL2}
A.A. Kordyuk et al., Appl. Phys. Lett. to be published;
LANL E-Print Archive, cond-mat/9905032.
\bibitem{PhyC}
A.A. Kordyuk et al., Physica C {\bf 310}, 173 (1998).
\bibitem{Clem1}
J.R. Clem, in LT13 Proc. {\bf 3}, 102 (1974).
\bibitem{Clem2}
J.R. Clem, J. Appl. Phys. {\bf 50}, 3518 (1979).
\bibitem{Burl}
L. Burlachkov, Phys. Rev. B {\bf 47}, 8056 (1993).
\end{thebibliography}
\end{document}